\documentclass{layoutp}

\usepackage{relsize}
\usepackage{amsmath}
\usepackage{mathtools}
\usepackage{cite}
\usepackage{float}

\newcommand{\sups}[1]{$^{\mathrm{#1}}$}

\begin{document}

\title{The context-dependence of mutations: a linkage of formalisms}

\author{Frank J. Poelwijk\affil{1}{Green Center for Systems Biology, UT Southwestern Medical Center, 6001 Forest Park Road, Dallas, TX 75235, USA}\sups{1}, Vinod Krishna\affil{1}{Green Center for Systems Biology, UT Southwestern Medical Center, 6001 Forest Park Road, Dallas, TX 75235, USA}\sups{2},\and Rama Ranganathan\!\affil{2}{Departments of Biophysics and Pharmacology and Green Center for Systems Biology, UT Southwestern Medical Center, 6001 Forest Park Road, Dallas, TX 75235, USA}\sups{1}}

\contributor{\sups{1}To whom correspondence should be addressed. E-mail: poelwijk@gmail.com or rama.ranganathan@utsouthwestern.edu\\
\sups{2\,}Present address: Janssen Pharmaceuticals Research \& Development, 1400 McKean Road, Spring House, PA 19454}

\maketitle

\begin{article}
\begin{abstract}
{Defining the extent of epistasis -- the non-independence of the effects of mutations -- is essential for understanding the relationship of genotype, phenotype, and fitness in biological systems.  The applications cover many areas of biological research, including biochemistry, genomics, protein and systems engineering, medicine, and evolutionary biology.  However, the quantitative definitions of epistasis vary among fields, and its analysis beyond just pairwise effects remains obscure in general.  Here, we show that different definitions of epistasis are versions of a single mathematical formalism - the weighted Walsh-Hadamard transform.  We discuss that one of the definitions, the backgound-averaged epistasis, is the most informative when the goal is to uncover the general epistatic structure of a biological system, a description that can be rather different from the local epistatic structure of specific model systems.  Key issues are the choice of effective ensembles for averaging and to practically contend with the vast combinatorial complexity of mutations.  In this regard, we discuss possible approaches for optimally learning the epistatic structure of biological systems.}
\end{abstract}

There has been much recent interest in the prevalence of epistasis in the relationships between genotype, phenotype, and fitness in biological systems \cite{Biochemistry29_8509, NatRevGenet9_855,CurrOpinBiotechnol22_66,TrendsGenet27_323,Science328_469,PLoSGenet6_e1001162,PLoSGenet7_e1001301}. Epistasis here is defined as the non-independence (or context-dependence) of the effect of a mutation, which is a generalization of Bateson's original definition of epistasis as a genetic interaction in which a mutation 'masks' the effect of variation at another locus \cite{Science26_647}.  It is also in line with Fisher's broader definition of 'epistacy' \cite{TransRSocEdinb52_399}.
Epistasis limits our ability to predict the function of a system that harbors several mutations given knowledge of the effects of those mutations taken independently \cite{JMolBiol196_733,CurrOpinStructBiol6_3,JMolBiol238_133,JBiolChem272_701}, and makes these relationships increasingly more complex \cite{PNAS101_111,Science265_2037,PLoSGenet7_e1002180,PLoSGenet8_e1002551,PNAS102_1572,Nature464_1039}. From an evolutionary perspective, the presence of epistatic interactions may limit or entirely preclude trajectories of single-mutation steps towards peaks in the fitness landscape \cite{Genetics167_559, Evolution59_1165, Nature445_383, JTheorBiol272_141,PNAS106_12025,Evolution67_2762,Evolution67_3120,MolSystBiol6_429,Science340_1324,eLife2_e00631}. With regard to human health, epistasis complicates our understanding of the origin and progression of disease \cite{Cell145_30,Cell155_21,CurrOpinGenetDev23_602,NatureGenetics43_487,NeurobiolAging30_1333,NatChemBiol2_458,CurrOpinGenetDev23_678,NatureComm5_4828}. Thus, interest in the extent of epistatic interactions in biological systems has originated from the fields of protein biochemistry, protein engineering, medicine, systems biology, and evolutionary biology alike.

Originally epistasis was considered in the context of two genes, but we can define it more broadly as the non-independence of mutational effects in the genome, whether the effects are within, between, or even outside protein coding regions (e.g. in regulatory regions). The perturbations may go beyond point mutagenesis, but we limit the discussion here for clarity of presentation. Importantly, the definition of epistasis can be extended beyond pairwise effects to comprise a hierarchy of 3-way, 4-way, and higher-order terms that represent the complete theoretical description of epistasis between the parts that make up a biological system.

How can we quantitatively assign an epistatic interaction given experimentally determined effects of mutations? Since epistasis is deviation from independence, it is crucial to first explicitly state the null hypothesis: asserting what exactly it means to have \emph{independent} contributions of mutations. This by itself can be non-trivial. In some cases the phenotype is directly related to a thermodynamic state variable, and the issue is then straightforward: independence implies additivity in the state variable. For example, for equilibrium binding reactions between two proteins, independence means additivity in the free energy of binding $\Delta{G_{bind}}$, such that the energetic effect of a double mutation is the sum of the energetic effects of each single mutation taken independently. However, in general, many phenotypes cannot be so directly linked to a thermodynamic state variable, and quantification of epistasis needs to be accompanied by a proper rationale for the choice of null hypothesis. In what follows we will assume this step has already been carried out and we will equate independence with additivity of mutational effects. Epistasis between two mutations is then defined as the degree to which the effect of both mutations together differs from the sum of the effects of the single mutations.

In this paper, we describe three theoretical frameworks that have been proposed for characterizing the epistasis between components of biological systems; these frameworks originate in different fields and use seemingly different calculations to describe the non-independence of mutations \cite{PNAS101_111,Biochemistry40_14012,JMolBiol267_696,JVirol81_12077,NatCellBiol8_571,NatureGenetics43_487,Science312_111,PNAS106_12025,ProteinEng14_633,PNAS107_9158,CurrOpinGenetDev23_700, JStatMech2013_P01005}. We show that these formalisms are different manifestations of a common mathematical principle, a finding that explains their conceptual similarities and distinctions. Each of these formalisms has its value depending on depth of coverage and nature of sampling in the experimental data, and the purpose of the analysis. In the end, the fundamental issue is to develop practical approaches for optimally learning the epistatic structure of biological systems in the face of explosive combinatorial complexity of possible epistatic interactions between mutations. Demonstrating the mathematical relationships between the different frameworks for analyzing epistasis is a first key step in this process.

\section{Results}
\subsection*{Basic definitions}\
\begin{figure}[t]
\centerline{\includegraphics[width=.45\textwidth]{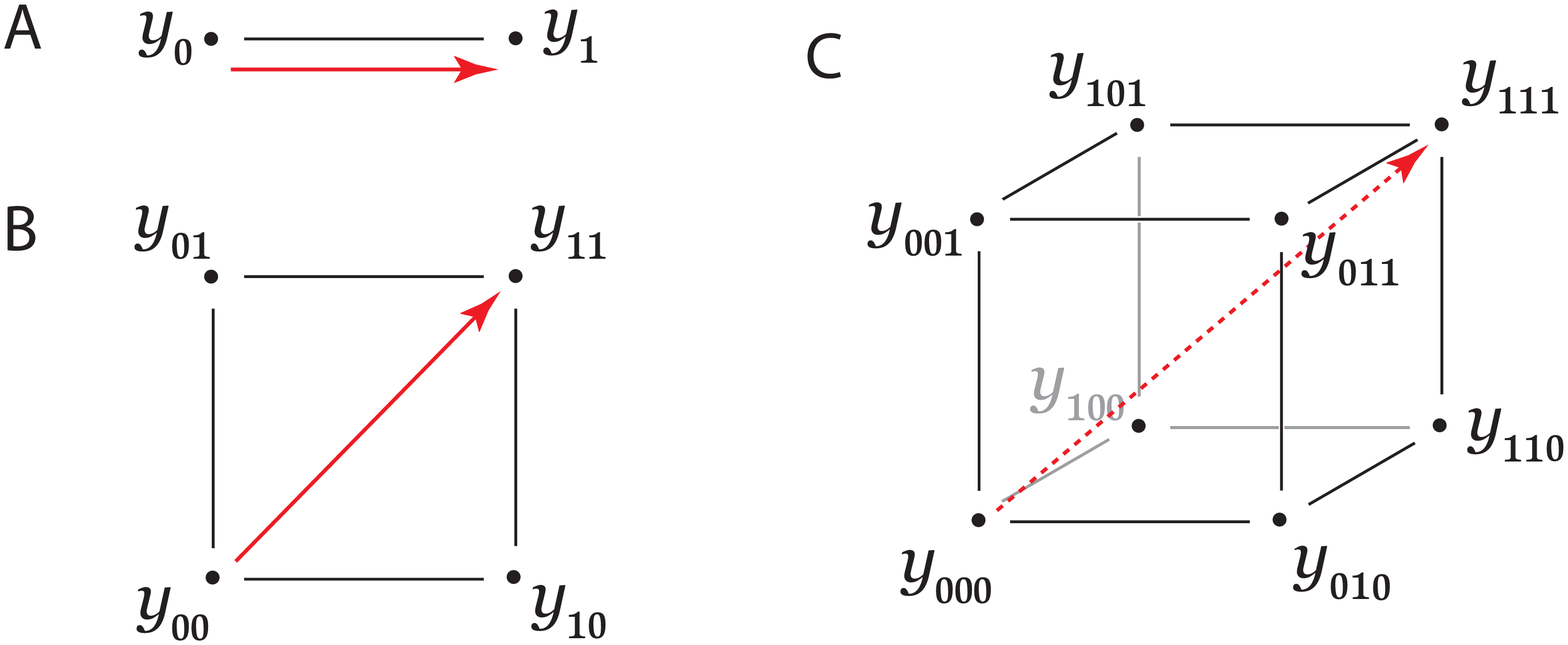}}
\vspace*{.3cm}
\caption{Representation of (A) single mutant, (B) double mutant, and (C) triple mutant experiments. Phenotypes are denoted by $y_g$, where $g$ is the underlying genotype. $g = \{g_N, ...,g_1\}$ with $g_i \in \{0,1\}$; '$0$' or '$1$' indicates the state of the mutable site (e.g., amino acid position). The effect of a single, double, triple mutation is given by the red arrows.  Pairwise (or second-order) epistasis is defined as the differential effect of a mutation depending on the background in which it occurs, for example in (B) it is the degree to which the effect of one mutation (e.g. $y_{10}-y_{00}$) deviates in the background of the second mutation ($y_{11}-y_{01}$).  Thus, the expression for second order epistasis is $(y_{11} - y_{10}) - (y_{01} - y_{00})$.  The third order and higher cases are considered in the main text,
\label{fig:explain}}
\end{figure}
We begin with a formal definition of genotype, phenotype, and the representation of mutational effects.  Consider a specific sequence comprised of $N$ positions as a binary string $g = \{g_N, ...,g_1\}$ with $g_i \in \{0,1\}$, where '$0$' and '$1$' represent the "wild-type" and mutant state of each position, respectively.  This defines a total space of $2^N$ genotypes.  The analysis could be expanded to the case of multiple substitutions per position, but we consider just the binary case for clarity here.  Each genotype $g$ has an associated phenotype $y_g$, which is of the form that the independent action of two mutations means additivity in $y$. For notational simplicity, we will simply write the genotype in a $k$-bit binary form, where $k$ is the order of the mutations that are considered. For example, the effect of a single mutation is simply $y_1 - y_0$, the difference in the phenotype between the mutant and wild-type states (Fig. \ref{fig:explain}A). The effect of a double mutant is given by $y_{11} - y_{00}$ (red arrow, Fig. \ref{fig:explain}B), and its linkage through paths of single mutations is defined by a two-dimensional graph (a square network) with four total genotypes.  Similarly, a triple mutant effect is $y_{111}-y_{000}$ (red arrow, Fig. \ref{fig:explain}C), and its linkage through paths of single mutations are enumerated on a three-dimensional graph (a cube) with eight total genotypes.  More generally, and as described by Horowitz and Fersht~\cite{JMolBiol214_613}, the phenotypic effect of any arbitrary $n$-dimensional mutation can be represented by an $n$-dimensional graph, with $2^n$ total genotypes. Understanding the relationship of the phenotypes of multiple mutants to that of the underlying lower-order mutant states is the essence of epistasis, and is described below.

\subsection*{The biochemical view of epistasis}\
A well-known approach in biochemistry for analyzing the cooperativity of amino acids in specifying protein structure and function is to use the formalism of thermodynamic mutant-cycles~\cite{JMolBiol196_733,JMolBiol214_613,FoldDes1_R121,JMolBiol224_733}, one manifestation of the general principle of epistasis.  In this approach, the "phenotype" is typically an equilibrium free energy $\Delta{G}$ (e.g. of thermodynamic stability or biochemical activity), and the goal is to obtain information about the structural basis of this phenotype through mutations that represent subtle perturbations of the wild-type state. For pairs of mutations, the analysis involves measurements of four variants: wild-type ($y_{00}=\Delta{G}^{\raisebox{1pt}{$\scriptstyle \mathrm{\,o}\!$}}_{0}\,$), each single mutant ($y_{01}=\Delta{G}^{\raisebox{1pt}{$\scriptstyle \mathrm{\,o}\!$}}_{1}$ and $y_{10}=\Delta{G}^{\raisebox{1pt}{$\scriptstyle \mathrm{\,o}\!$}}_{2}\,$), and the double mutant ($y_{11}=\Delta{G}^{\raisebox{1pt}{$\scriptstyle \mathrm{\,o}\!$}}_{1,2}$), where the subscripts designate the mutated positions, and the superscript 'o' indicates free energy relative to a standard state (Fig. 1B).

From this, we can compute a coupling free energy between the two mutations ($\Delta^{2}G_{1,2}$) as the degree to which the effect of one mutation ($\Delta^{\raisebox{3pt}{$\scriptstyle \!1\!$}}{G}_{1}$) is different when tried in the background of the other mutation ($\Delta^{\raisebox{3pt}{$\scriptstyle \!1\!$}}{G}_{1|2}$):
\begin{align}
\Delta^{\raisebox{3pt}{$\scriptstyle \!2\!$}}{G}_{1,2} &= \Delta^{\raisebox{3pt}{$\scriptstyle \!1\!$}}{G}_{1}|_2 - \Delta^{\raisebox{3pt}{$\scriptstyle \!1\!$}}{G}_{1} \nonumber \\ &= (\Delta{G}^{\raisebox{1pt}{$\scriptstyle \mathrm{\,o}\!$}}_{1,2} - \Delta{G}^{\raisebox{1pt}{$\scriptstyle \mathrm{\,o}\!$}}_{2}\,) - (\Delta{G}^{\raisebox{1pt}{$\scriptstyle \mathrm{\,o}\!$}}_{1} - \Delta{G}^{\raisebox{1pt}{$\scriptstyle \mathrm{\,o}\!$}}_{0}\,)
\label{eq:pair_energy}
\end{align}
Whereas the $\Delta{G}^{\raisebox{1pt}{$\scriptstyle \mathrm{\,o}\!$}}$ terms are individual measurements and $\Delta^{\raisebox{3pt}{$\scriptstyle \!1\!$}}{G}$ terms are the effects of single mutations relative to wild-type, $\Delta^{\raisebox{2pt}{$\scriptstyle \!2\!$}}{G}$ is a second order epistatic term describing the cooperativity (or non-independence) of two mutations with respect to the wild-type state.  This analysis can be expanded to higher order.  For example, the third order epistatic term describing the cooperative action of three mutations 1, 2, and 3 ($\Delta^{\raisebox{3pt}{$\scriptstyle \!3\!$}}{G}_{1,2,3} $) is defined as the degree to which the second order epistasis of any two mutations is different in the background of the third mutation:
\begin{align}
\Delta^{\raisebox{3pt}{$\scriptstyle \!3\!$}}{G}_{1,2,3} &= \Delta^{\raisebox{3pt}{$\scriptstyle \!2\!$}}{G}_{1,2}|_3 - \Delta^{\raisebox{3pt}{$\scriptstyle \!2\!$}}{G}_{1,2} \nonumber \\ &= \Delta{G}^{\raisebox{1pt}{$\scriptstyle \mathrm{\,o}\!$}}_{1,2,3} - \sum\limits^{\mathsmaller{3}}_{\mathsmaller{i<j}}\Delta{G}^{\raisebox{1pt}{$\scriptstyle \mathrm{\,o}\!$}}_{i,j} +
\sum\limits^{\mathsmaller{3}}_{\mathsmaller{i}}  \Delta{G}^{\raisebox{1pt}{$\scriptstyle \mathrm{\,o}\!$}}_{i} - \Delta{G}^{\raisebox{1pt}{$\scriptstyle \mathrm{\,o}\!$}}_{0}
\label{eq:triple_energy}
\end{align}
Note that $\Delta^{\raisebox{3pt}{$\scriptstyle \!3\!$}}{G}$ requires measurement of eight individual genotypes (Fig. 1C).  More generally, we can define an $n$-th order epistatic term ($\Delta^{\raisebox{2pt}{$\scriptstyle \!n\!$}}{G}$), describing the cooperativity of $n$ mutations,
\begin{align}
\Delta^{\raisebox{3pt}{$\scriptstyle \!n\!$}}{G}_{\mathsmaller{\mathsmaller{ 1,\ldots, n}}} &= \Delta{G}_{\mathsmaller{\scriptsize 1,\ldots, n}}^{\raisebox{1pt}{$\scriptstyle \mathrm{\,o}$}} \,+\, \mathsmaller{\mathsmaller{(-1)^1}} \hspace{-.7cm} \sum\limits^{n}_{i_{\mathsmaller{1}}\mathsmaller{<}i_{\mathsmaller{2}}\mathsmaller{<\ldots<}i_{\mathsmaller{n-\mathsmaller{1}}}} \hspace{-.6cm} \Delta{G}_{i_\mathsmaller{1},i_\mathsmaller{2},\mathsmaller{\ldots}, i_\mathsmaller{n-\mathsmaller{1}}}^{\raisebox{1pt}{$\scriptstyle \mathrm{\,o}$}} \,\nonumber\\ &+ \,
\mathsmaller{\mathsmaller{(-1)^2}} \hspace{-.7cm} \sum\limits^{n}_{i_{\mathsmaller{1}}\mathsmaller{<}i_{\mathsmaller{2}}\mathsmaller{<\ldots<}i_{\mathsmaller{n-\mathsmaller{2}}}} \hspace{-.6cm}\Delta{G}_{i_\mathsmaller{1},i_\mathsmaller{2},\mathsmaller{\ldots}, i_\mathsmaller{n-\mathsmaller{2}}}^{\raisebox{1pt}{$\scriptstyle \mathrm{\,o}$}} \,+ \ldots +\,  \mathsmaller{\mathsmaller{(-1)^{n}}} \Delta{G}_{0}^{\raisebox{1pt}{$\scriptstyle \mathrm{\,o}$}}
\label{eq:expansion}
\end{align}
It is possible to write this expansion in a compact matrix form:
\begin{equation}
\boldsymbol{\bar{\gamma}} = \boldsymbol{G} \boldsymbol{\bar{y}}
\end{equation}
where $\boldsymbol{\bar{\gamma}}$ is the vector of $2^n$ epistasis terms of all orders, and $\boldsymbol{\bar{y}}$ is the vector of $2^n$ free energies corresponding to phenotypes of all the individual variants listed in binary order.  To illustrate, for three mutations $n = 3$, and we obtain
\begin{equation*}
\begin{pmatrix}
\gamma_{000}\\
\gamma_{001}\\
\gamma_{010}\\
\gamma_{011}\\
\gamma_{100}\\
\gamma_{101}\\
\gamma_{110}\\
\gamma_{111}
\end{pmatrix}
=
\setlength{\arraycolsep}{2pt}
\begin{pmatrix*}[r]
1 & 0 & 0 & 0 & 0 & 0 & 0 &\ \ 0\ \\
-1 & 1 & 0 & 0 & 0 & 0 & 0 &\ \ 0\ \\
-1 & 0 & 1 & 0 & 0 & 0 & 0 &\ \ 0\ \\
1 & -1 & -1 & 1 & 0 & 0 & 0 &\ \ 0\ \\
-1 & 0 & 0 & 0 & 1 & 0 & 0 &\ \ 0\ \\
1 & -1 & 0 & 0 & -1 & 1 & 0 &\ \ 0\ \\
1 & 0 & -1 & 0 & -1 & 0 & 1 &\ \ 0\ \\
-1 & 1 & 1 & -1 & 1 & -1 & -1 &\ \ 1\
\end{pmatrix*}
\setlength{\arraycolsep}{6pt}
*
\begin{pmatrix*}
y_{000}\\
y_{001}\\
y_{010}\\
y_{011}\\
y_{100}\\
y_{101}\\
y_{110}\\
y_{111}
\end{pmatrix*}
\end{equation*}
In this representation, subscripts in $\boldsymbol{\bar{y}}$ represent combinations of mutations (e.g. ${y}_{011} = \Delta{G}^{\raisebox{1pt}{$\scriptstyle \mathrm{\,o}\!$}}_{1,2}$, a double mutant) and subscripts in $\boldsymbol{\bar{\gamma}}$ represent epistatic order (e.g. $\gamma_{011} = \Delta^{\raisebox{3pt}{$\scriptstyle \!2\!$}}{G}_{1,2} $, pairwise epistasis between mutations 1 and 2). Thus, equations \refeq{eq:pair_energy} and \refeq{eq:triple_energy} correspond to multiplying $\boldsymbol{\bar{y}}$ by the fourth or eighth row of $\boldsymbol{G}$, respectively, to specify $\gamma_{011}$ and $\gamma_{111}$.  Note that $\boldsymbol{\bar{y}}$ and $\boldsymbol{\bar{\gamma}}$ contain precisely the same information, re-written in a different form.  The matrix $\boldsymbol{G}$ represents an operator linking these two representations of the mutation data and we will return to the nature of the operation in a later section.  We can write a recursive definition for $\boldsymbol{G}$ that defines the mapping between $\boldsymbol{\bar{y}}$ and $\boldsymbol{\bar{\gamma}}$ for all epistatic orders $n$:
\begin{equation}
\boldsymbol{G}_{n+1}
=
\setlength{\arraycolsep}{2pt}
\begin{pmatrix*}[r]
 \boldsymbol{G}_n & \ \ 0 \ \   \\[0.2em]
 -\boldsymbol{G}_n & \boldsymbol{G}_n
\end{pmatrix*}\ \ \mathrm{with}\ \ \
\setlength{\arraycolsep}{6pt}
\boldsymbol{G}_0 = 1
\label{eq:Grecursive}
\end{equation}
The inverse mapping is defined by $\boldsymbol{\bar{y}} = \boldsymbol{G}^{-1} \boldsymbol{\bar{\gamma}}$.  This relationship gives the effect of any combination of mutants (in $\boldsymbol{\bar{y}}$) as a sum over epistatic terms (in $\boldsymbol{\bar{\gamma}}$). For example, the energetic effect of three mutations 1,2, and 3 ($\Delta{G}^{\raisebox{1pt}{$\scriptstyle \mathrm{\,o}\!$}}_{1,2,3} = y_{111}$) is:
\begin{align}
\Delta{G}^{\raisebox{1pt}{$\scriptstyle \mathrm{\,o}\!$}}_{1,2,3} = \Delta^{\raisebox{3pt}{$\scriptstyle \!3\!$}}{G}_{1,2,3} + \sum\limits^{\mathsmaller{3}}_{\mathsmaller{i<j}}\Delta^{\raisebox{3pt}{$\scriptstyle \!2\!$}}{G}_{i,j} + \sum\limits^{\mathsmaller{3}}_{\mathsmaller{i}}\Delta^{\raisebox{3pt}{$\scriptstyle \!1\!$}}{G}_{i} + \Delta{G}^{\raisebox{1pt}{$\scriptstyle \mathrm{\,o}\!$}}_{0}
\end{align}
Thus, in the most general case, the free energy value of a multiple mutation requires knowledge of the effect of the single mutations and all associated epistatic terms.  For the triple mutant, this means the wild-type phenotype, the three single mutant effects, the three two-way epistatic interactions, and the single three-way epistatic term.  This analysis highlights two important properties of epistasis: (1) the lack of any epistatic interactions between mutations dramatically simplifies the description of multiple mutations to just the sum over the underlying single mutation effects, and (2) the absence of lower-order epistatic interactions (e.g. $\Delta^{\raisebox{3pt}{$\scriptstyle \!2\!$}}{G}_{i,j} = 0$) does not imply absence of higher order epistatic terms.

\subsection*{The ensemble view of epistasis}\
In contrast to the biochemical definition, the significance of a mutation (and its epistatic interactions) may also be defined not solely with regard to a single reference state as the "wild-type", but as an average over many possible genotypes.  As we show below, such averaging better represents the epistatic level at which mutations operate, and in principle, can separate mutant effects that are idiosyncratic to particular genotypes from those that are fundamentally important.  The concept of averaging epistasis over genotypic backgrounds is analogous to the idea of the 'schema average fitness' in the field of genetic algorithms (GA) \cite{ComplexSystems3_129,AmJPhys49_466}, which was recently introduced in biology \cite{CurrOpinGenetDev23_700}.

In its complete form, background-averaged epistasis considers averages over all possible genotypes for the remaining positions in the ensemble. For example, if $n=3$, the epistasis between two positions 1 and 2 is computed as an average over both states of the third position ($\varepsilon_{*11}$, with the averaging denoted by '$*$') (see. Fig. 1C):
\begin{multline}
\varepsilon_{*11} = \frac{1}{2} \bigg\{ [(y_{111} - y_{110}) - (y_{101} - y_{100})]\\
+ [(y_{011} - y_{010}) - (y_{001} - y_{000})] \bigg\}
\end{multline}
Thus for $n=3$, we can write all epistatic terms:
\begin{equation*}
\begin{pmatrix}
\varepsilon_{***}\\
\varepsilon_{**1}\\
\varepsilon_{*1*}\\
\varepsilon_{*11}\\
\varepsilon_{1**}\\
\varepsilon_{1*1}\\
\varepsilon_{11*}\\
\varepsilon_{111}
\end{pmatrix}
=
\boldsymbol{V} *
\setlength{\arraycolsep}{2pt}
\begin{pmatrix*}[r]
\ 1 & 1 & 1 & 1 & 1 & 1 & 1 & 1\ \\
\ 1 & -1 & 1 & -1 & 1 & -1 & 1 & -1\ \\
\ 1 & 1 & -1 & -1 & 1 & 1 & -1 & -1\ \\
\ 1 & -1 & -1 & 1 & 1 & -1 & -1 & 1\ \\
\ 1 & 1 & 1 & 1 & -1 & -1 & -1 & -1\ \\
\ 1 & -1 & 1 & -1 & -1 & 1 & -1 & 1\ \\
\ 1 & 1 & -1 & -1 & -1 & -1 & 1 & 1\ \\
\ 1 & -1 & -1 & 1 & -1 & 1 & 1 & -1\
\end{pmatrix*}
\setlength{\arraycolsep}{6pt}
*
\begin{pmatrix*}
y_{000}\\
y_{001}\\
y_{010}\\
y_{011}\\
y_{100}\\
y_{101}\\
y_{110}\\
y_{111}
\end{pmatrix*}
\end{equation*}
where $\boldsymbol{V}$ is a diagonal weighting matrix to account for averaging over different number of terms as a function of the order of epistasis; $v_{ii} = (-1)^{q_i}/2^{n-q_i}$, where $q_i$ is the order of the epistatic contribution in row $i$.  More generally, for any number of mutations $n$:
\begin{equation}
\boldsymbol{\bar{\varepsilon}} = \boldsymbol{V} \boldsymbol{H} \boldsymbol{\bar{y}}.
\end{equation}
where $\boldsymbol{\bar{y}}$ is the same vector of phenotypes of variants as defined above, $\boldsymbol{\bar{\varepsilon}}$ is the vector of background averaged epistatic terms, and $\boldsymbol{H}$ is the operator for background-averaged epistasis, defined recursively as
\begin{equation}
\boldsymbol{H}_{n+1}
=
\setlength{\arraycolsep}{2pt}
\begin{pmatrix*}[r]
\ \boldsymbol{H}_n & \boldsymbol{H}_n\\[0.2em]
\ \boldsymbol{H}_n & -\boldsymbol{H}_n
\end{pmatrix*}\ \ \mathrm{with}\ \ \
\setlength{\arraycolsep}{6pt}
\boldsymbol{H}_0 = 1
\end{equation}
The recursive definition for the weighting matrix $\boldsymbol{V}$ is
\begin{equation}
\boldsymbol{V}_{n+1}
=
\setlength{\arraycolsep}{2pt}
\begin{pmatrix*}[c]
\ \frac{1}{2}\boldsymbol{V}_n & 0 \\[0.2em]
\  0& -\boldsymbol{V}_n\
\end{pmatrix*}\ \ \mathrm{with}\ \ \
\setlength{\arraycolsep}{6pt}
\boldsymbol{V}_0 = 1
\end{equation}

The matrix $\boldsymbol{H}$ has special significance; its action mathematically corresponds to a generalized Fourier analysis~\cite{JAmStatAssoc86_461} known as the Walsh-Hadamard transform. This converts the phenotypes of individual variants (in $\boldsymbol{\bar{y}}$) into a vector of averaged epistasis (in $\boldsymbol{\bar{\varepsilon}}$), an operation that can also be seen as a spectral analysis of the high-dimensional phenotypic landscape defined by the genotypes studied.  In this transform, the phenotypic effects of combinations of mutations are represented as sums over averaged epistatic terms.

In summary, the definition of epistasis proposed in evolutionary genetics is a global definition over sequence space, averaging the epistatic effects of mutations over the ensemble of all possible variants. In contrast, the biochemical definition given in the previous section is a local one, treating a particular variant as a reference for determining the epistatic effect of mutations.

\subsection*{Estimating epistasis with linear regression}\
A third approach for analyzing epistasis is linear regression. For example, when we have a complete dataset of phenotypes of all $2^n$ genotypes, we can use regression to define the extent to which epistasis is captured by only considering terms to some order $r < n$.  That is, whether terms up to the $r$\sups{th} order are sufficient for effectively capturing the full complexity of a biological system. The standard form for a linear regression is a set of equations:
\begin{equation}
y_g = \beta_0 + \sum^n_{i=1} \beta_i g_i + \sum^n_{i<j} \beta_{ij} g_i g_j + \sum^n_{i<j<k} \beta_{ijk} g_i g_j g_k + ... + \epsilon_g
\end{equation}
for each genotype $g$. The $\beta$ terms denote the regression coefficients corresponding to the (epistatic) effects between subscripted positions, and $\epsilon_g$ is the residual noise term.
In matrix form this can be written as
\begin{equation}
\boldsymbol{\bar{y}} = \boldsymbol{X} \boldsymbol{\bar{\beta}} + \boldsymbol{\bar{\epsilon}}.
\end{equation}
where $\boldsymbol{X}$ tabulates which regression coefficients are summed over for genotypes $g$. For $n = 3$, regressing to full order, we can write
\begin{equation*}
\begin{pmatrix*}
y_{000}\\
y_{001}\\
y_{010}\\
y_{011}\\
y_{100}\\
y_{101}\\
y_{110}\\
y_{111}
\end{pmatrix*}
=
\begin{pmatrix*}[r]
1 & 0 & 0 & 0 & 0 & 0 & 0 & 0\\
1 & 1 & 0 & 0 & 0 & 0 & 0 & 0\\
1 & 0 & 1 & 0 & 0 & 0 & 0 & 0\\
1 & 1 & 1 & 1 & 0 & 0 & 0 & 0\\
1 & 0 & 0 & 0 & 1 & 0 & 0 & 0\\
1 & 1 & 0 & 0 & 1 & 1 & 0 & 0\\
1 & 0 & 1 & 0 & 1 & 0 & 1 & 0\\
1 & 1 & 1 & 1 & 1 & 1 & 1 & 1
\end{pmatrix*}
*
\begin{pmatrix}
\beta_{000}\\
\beta_{001}\\
\beta_{010}\\
\beta_{011}\\
\beta_{100}\\
\beta_{101}\\
\beta_{110}\\
\beta_{111}
\end{pmatrix}
+ \boldsymbol{\bar{\epsilon}}
\end{equation*}
following the same rule for subscripts as before.
$\boldsymbol{X}$ has the recursive definition:
\begin{equation}
\boldsymbol{X}_{n+1}
=
\setlength{\arraycolsep}{2pt}
\begin{pmatrix*}[r]
\ \boldsymbol{X}_n & \ \ 0 \ \ \    \\[0.2em]
\ \boldsymbol{X}_n & \boldsymbol{X}_n\
\end{pmatrix*}\ \ \mathrm{with}\ \ \
\setlength{\arraycolsep}{6pt}
\boldsymbol{X}_0 = 1
\label{eq:recursiveX}
\end{equation}
It is worth noting that the inverse of $\boldsymbol{X}$ is $\boldsymbol{X}^{-1} = \boldsymbol{G}$, the operator for biochemical epistasis (Eq. \refeq{eq:Grecursive}; see Supplementary Information). Thus, the multi-dimensional mutant-cycle analysis is indistinguishable from regression to full order -- the case in which $r = n$ and $\boldsymbol{\bar{\epsilon}} = 0$.

However, the usual aim of regression is to approximate the data with fewer coefficients than there are data points, i.e., $r < n$. To express this, we simply remove the columns from $\boldsymbol{X}$ that refer to the epistatic orders excluded from the regression (i.e., $> r$): $\boldsymbol{X}$ is multiplied by an $2^n$-by-$m$ matrix $\boldsymbol{Q}$, the identity matrix with columns corresponding to epistatic orders higher than $r$ removed. $m$ is the number of epistatic terms up to $r$ and is given by $m = \sum^{r}_{i=0} {n \choose i}$. Thus for regression to order $r$, we can define $\boldsymbol{\hat{X}} = \boldsymbol{X} \boldsymbol{Q}$, and write
\begin{equation}
\boldsymbol{\bar{y}} = \boldsymbol{\hat{X}} \boldsymbol{\hat{\beta}} + \boldsymbol{\hat{\epsilon}}.
\end{equation}
The linear regression is performed by solving the so-called normal equations
\begin{equation}
\boldsymbol{\hat{\beta}} = (\boldsymbol{\hat{X}}^T \boldsymbol{\hat{X}})^{-1} \boldsymbol{\hat{X}}^T  \boldsymbol{\bar{y}}
\label{eq:NormalEq}
\end{equation}
where $\boldsymbol{\hat{X}}^T \boldsymbol{\hat{X}}$ is necessarily square and invertible as long as $\boldsymbol{\hat{X}}$ is full column rank and hence $\boldsymbol{\hat{X}}^T \boldsymbol{\hat{X}}$ is full rank.
Note that in this analysis we compute epistatic terms only up to the $r$\sups{th} order, but use phenotype/fitness data of all $2^n$ combinations of mutants. The more general case in which we estimate epistatic terms with less than $2^n$ data points is distinct and is discussed below.

If the biochemical definition of epistasis is a local one, exploring the coupling of mutations of all order with regard to one "wild-type" reference, and the ensemble view of epistasis is a global one, assessing the coupling of mutations of all order averaged over all possible genotypes, then the regression view of epistasis is an attempt to project to a lower dimension - capturing epistasis as much as possible with low-order terms.

\subsection*{Link between the formalisms}\
The analysis presented above leads to a simple unifying concept underlying the calculations of epistasis. In general, all the calculations are a mapping from the space of phenotypic measurements of genotypes $\boldsymbol{\bar{y}}$ to epistatic coefficients $\boldsymbol{\bar{\omega}}$, in a general form $\boldsymbol{\bar{\omega}} = \boldsymbol{\Omega}_{epi}\,\boldsymbol{\bar{y}}$, where $\boldsymbol{\Omega}_{epi}$ is the epistasis operator. We give the bottom line of the different operators below; their formal mathematical derivations can be found in the Supplementary Information.

The most general situation is that of the background-averaged epistasis with averaging over the complete space of possible genotypes. In this case
\begin{equation}
\boldsymbol{\Omega}_{epi} = \boldsymbol{V} \boldsymbol{H}, \label{eq:OmegaHadamard}
\end{equation}
where $\boldsymbol{H}$ is a $2^n \times 2^n$ matrix corresponding to the Walsh-Hadamard transform ($n$ is the number of mutated sites) and $\boldsymbol{V}$ is a matrix of weights to normalize for the different numbers of terms for epistasis of different orders.  The biochemical definition of epistasis using one "wild-type" sequence as a reference is a sub-sampling of terms in the Hadamard transform.  In this case
\begin{equation}
\boldsymbol{\Omega}_{epi} = \boldsymbol{V} \boldsymbol{X}^T \boldsymbol{H}, \label{eq:DeltaGHadamard}
\end{equation}
where $\boldsymbol{X}$ is, as defined in Eq. \refeq{eq:recursiveX}. In essence, $\boldsymbol{X}^T$ picks out the terms in $\boldsymbol{H}$ that concern the wild-type background.  Note that both these mapping are one-to-one, such that the number of epistatic terms (in $\boldsymbol{\bar{\omega}}$) is equal to the number of phenotypic measurements (in $\boldsymbol{\bar{y}}$) and no information is lost. In contrast, regression to lower orders necessarily implies fewer epistatic terms than data points, which means the mapping is compressive and information is lost. In this case
\begin{equation}
\boldsymbol{\Omega}_{epi} = \boldsymbol{V} \boldsymbol{X}^T \boldsymbol{S} \boldsymbol{H}, \label{eq:regressHadamard}
\end{equation}
where $\boldsymbol{S}$ ($\equiv \boldsymbol{Q} \boldsymbol{Q}^T$) is the identity matrix but with zeros on the diagonal at the orders that are higher than which we regress over.

The fundamental point is that all three formalisms for computing epistasis are just versions of the Walsh-Hadamard transform, with terms selected as appropriate for the choice of a single reference sequence or limitations on the order of epistatic terms considered. From a computational point of view, it is interesting to note that regression using the Hadamard transform makes matrix inversion unnecessary (compare with Eq. \refeq{eq:NormalEq}).

\subsection*{An empirical example: a cooperative mechanism in a PDZ domain}
\begin{figure}[t]
\centerline{\includegraphics[width=.47\textwidth]{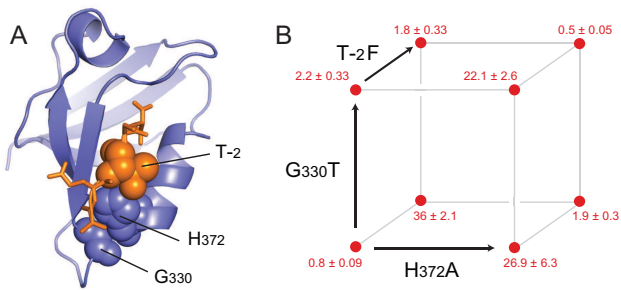}}
\vspace*{.2cm}
\caption{Example of three-way epistasis in the affinity of a PDZ binding domain for its ligand. A) In blue the PSD95-PDZ3 domain, and in orange its ligand peptide positioned in the binding pocket. The positions under consideration are shown as space-filling spheres. B) Measured $K_d$ values in $\mu$M for all eight combinations of two amino acids at the three mutable positions.\label{fig:PDZ}}
\end{figure}

\
To illustrate the different analyses of epistasis, we consider a small case study of three spatially proximal mutations that define a switch in ligand specificity in PSD95-PDZ3, a member of the PDZ family of protein interaction modules (Fig. \ref{fig:PDZ}A).  Two mutations are in PSD95-PDZ3 (G330T and H372A), and one mutation in its cognate ligand peptide (T-2F). The phenotype is the binding affinity, $K_d$, and the absence of epistasis implies additivity in the corresponding free energy, expressed as $\Delta{G}^{\raisebox{1pt}{$\scriptstyle \mathrm{o}\!$}} = RT \mathrm{ln} K_d$ in kcal mol\sups{-1}. Binding affinities for this system are from ref. \cite{Nature491_138}, and given in Figure \ref{fig:PDZ}B. These quantitative phenotypes are then transformed to epistatic terms using Eq.\,\ref{eq:OmegaHadamard}-\ref{eq:regressHadamard} (Table 1).

A number of simple mathematical relationships are evident in the data.  First, regression is carried out only to the second-order and therefore the third-order epistatic term for this analysis does not exist (or, equivalently, is set to zero if the epistatic vector $\boldsymbol{\hat{\beta}}$ is defined to be of full length $2^n$). Second, there are some equalities. The regression terms at the highest order (second, in this case) are equal to the corresponding terms for the averaged epistasis. This is because $\boldsymbol{X}^T \boldsymbol{S}$ sets columns corresponding to orders higher than the regression order to zero, leaving rows corresponding to the highest regression order with only one non-zero element, on the diagonal. For these rows the entries in the epistasis operators $\boldsymbol{V} \boldsymbol{X}^T \boldsymbol{S} \boldsymbol{H}$ and $\boldsymbol{V} \boldsymbol{H}$ are equal. Another more trivial equality is the highest-order term for the mutant-cycle and averaged epistasis formalisms; there is only one contribution for the highest order, and therefore no backgrounds to average over.

\begin{table}[b]
\caption{Interaction terms after applying the three different transforms to the PDZ-ligand dataset with three mutable positions: three-way mutant-cycle, background-averaged epistasis, and regression (to second order).}
\begin{tabular}{@{\vrule height 9.5pt depth4pt  width0pt}rr|rrrrr}
\hline
\footnotesize{genotype}\tablenote{The three mutable positions in genotypes are T-2F in the ligand, and H372A and G330T in the protein, respectively. They are designated in this column as 'THG'.}  & \footnotesize{free\ \ \ } & \footnotesize{\ interaction }& \footnotesize{mutant\ \ }& \footnotesize{bg. ave.\ }&\footnotesize{regression}\\[-.7em]

\footnotesize{\smaller{THG}\, \, \ \  }& \footnotesize{energy}\tablenote{Measured free energies in kcal/mol, expressed as $RT \mathrm{ln} K_d$, at $T$ = 293K} & \footnotesize{term}\tablenote{Interacting positions are in the same order as genotypes, for example '*11' indicates the epistasis between amino acid positions 372 and 330 in PSD95-PDZ3.}\ \ \ \ & \footnotesize{cycle\ \ \ } & \footnotesize{epistasis\ }& \footnotesize{terms\ \ \ \ }\\[-.4em]

& \footnotesize{$\boldsymbol{\bar{y}}$\ \ \ \ } & & \footnotesize{$\boldsymbol{\bar{\gamma}}$\ \ \ \ \ } & \footnotesize{$\boldsymbol{\bar{\varepsilon}}$\ \ \ \ \ }& \footnotesize{$\boldsymbol{\hat{\beta}}$\ \ \ \ \ \ }\\[.5em]

\footnotesize{000\ \ \ \ \  }& \footnotesize{$-$8.17\ \ }   & \footnotesize{***\ \ \ \ \ }  &	\footnotesize{$-$8.17\ \ \ }	&	\footnotesize{$-$7.24\ \ }    &	\footnotesize{$-$7.96\ \ \ }  \\[-.6em]
\footnotesize{001\ \ \ \ \  }& \footnotesize{$-$7.58\ \ }   & \footnotesize{**1\ \ \ \ \ }	&	\footnotesize{0.59\ \ \ }  	    &	\footnotesize{$-$0.51\ \ }	&	\footnotesize{0.17\ \ \ }	    \\[-.6em]
\footnotesize{010\ \ \ \ \  }& \footnotesize{$-$6.13\ \ }   & \footnotesize{*1*\ \ \ \ \ }	&	\footnotesize{2.05\ \ \ }    	&	\footnotesize{0.23\ \ }  	    &	\footnotesize{1.63\ \ \ }	    \\[-.6em]
\footnotesize{011\ \ \ \ \  }& \footnotesize{$-$6.24\ \ }   & \footnotesize{*11\ \ \ \ \ }	&	\footnotesize{$-$0.70\ \ \ }	&	\footnotesize{0.13\ \ }       &	\footnotesize{0.13\ \ \ }	    \\[-.6em]
\footnotesize{100\ \ \ \ \  }& \footnotesize{$-$5.96\ \ }   & \footnotesize{1**\ \ \ \ \ }	&	\footnotesize{2.22\ \ \ }  	    &	\footnotesize{$-$0.41\ \ }	&	\footnotesize{1.80\ \ \ }     \\[-.6em]
\footnotesize{101\ \ \ \ \  }& \footnotesize{$-$7.70\ \ }   & \footnotesize{1*1\ \ \ \ \ }	&	\footnotesize{$-$2.33\ \ \ }	&	\footnotesize{$-$1.50\ \ }	&	\footnotesize{$-$1.50\ \ \ }	\\[-.6em]
\footnotesize{110\ \ \ \ \  }& \footnotesize{$-$7.67\ \ }   & \footnotesize{11*\ \ \ \ \ }	&	\footnotesize{$-$3.76\ \ \ }	&	\footnotesize{$-$2.92\ \ }	&	\footnotesize{$-$2.92\ \ \ }	\\[-.6em]
\footnotesize{111\ \ \ \ \  }& \footnotesize{$-$8.45\ \ }   & \footnotesize{111\ \ \ \ \ }	&	\footnotesize{1.67\ \ \ }  	    &	\footnotesize{1.67\ \ }  	    &	\footnotesize{$0$\, \ \ \ }	    \\[-1em]
&&&&&&\\
\hline
\end{tabular}\ \\
\label{table:PDZ}
\end{table}

The data also illustrate the key properties of the different formalisms. The G330T, H372A, and T-2F mutations represent a collectively cooperative set of perturbations, as indicated by a significant third-order epistatic term by both mutant cycle and background averaged definitions ($\gamma_{111}=\varepsilon_{111}=1.67$ kcal mol\sups{-1}). But the three formalisms differ in the energetic value of the lower order epistatic terms. For example, G330T is essentially neutral for wild-type ligand binding but shows a dramatic gain in affinity in the context of the T-2F ligand; thus, a large second-order epistatic term by the biochemical definition ($\gamma_{101}=-2.33$ kcal mol\sups{-1}). However, the coupling between G330T and T-2F is nearly negligible in the background of H372A; as a consequence, the background averaged second-order epistasis term $\varepsilon_{1*1}$ is smaller ($-1.5$ kcal mol\sups{-1}). Similarly, both biochemical and regression formalisms assign a large first-order effect to the T-2F (1**) and H372A (*1*) single mutations, while the corresponding background-averaged terms are nearly insignificant.  For example, the free energy effect of mutating H372A ($\gamma_{010}$) is $2.05$ kcal mol\sups{-1} in the wild-type background, but is $-1.71$ kcal mol\sups{-1} in the background of the T-2F ligand mutation - a nearly complete reversal of the effect of this mutation depending on context.  Thus with background averaging, the first order term for H372A ($\varepsilon_{*1*}$) is close to zero. This makes sense; given the experiment described in Figure 2, the H372A mutation should not be thought of as a general determinant of ligand affinity.  Instead it is a conditional determinant, with an effect that depends on the identity of the amino acid at the $-2$ position of the ligand.  Note that the degree of averaging depends on the number of mutated sites, and thus the interpretation of mutational effects will depend on the scale of the experimental study.

These examples show that background averaging has the effect of "correcting" mutational effects for the existence of higher-order epistatic interactions. Without background averaging, the effect of a mutation (at any order) idiosyncratically depends on a particular reference genotype and will fail to account for higher order epistasis which modulates the observed mutational effect. Thus, background averaging provides a measure of the effects of mutation that represents its general value over many related systems, and more appropriately represents the cooperative unit within which the mutation operates.

\subsection*{The epistatic structure of real systems}\
The analytical expressions in Eq.\,\ref{eq:OmegaHadamard}-\ref{eq:regressHadamard} involves the measurement of phenotypes ($\boldsymbol{\bar{y}}$) for all $2^n$ combinatorial mutants, a fact that exposes two fundamental problems.  First, it is only practical when $n$ is small.  In such cases (e.g Figure 2, $n=3$), the data can be combinatorially complete permitting a full analysis - the local and global structure of epistasis, possible evolutionary trajectories, and adaptive trade-offs \cite{CurrOpinMicrobiol21_51}.  But for the typical size of protein domains ($n \sim 150$), the combinatorial complexity of mutations precludes the collection of complete datasets.  Second, even if it were possible, the sampling of all genotypes is not desired; indeed, the majority of systems in such an ensemble are unlikely to be functional and and averages over them are not meaningful with regard to learning the epistatic structure of native systems.  How then can we apply these epistasis formalisms in practice, especially with regard to background averaging?

To develop general principles, we begin with two obvious approaches that lead to well-defined alternative expressions for averaged epistasis. First, consider the case in which the data are only "locally complete"; that is, we have all possible mutants up to a certain order $p\le{}n$.  We can then define a measure that is intermediate between epistasis with a single reference genotype and epistasis with full background-averaging, which we will refer to as the \emph{partial} background-averaged epistasis.  For example, for three positions ($n=3$) with data complete only up to order ($p=2$), the partial background-averaged effect of the first position (rightmost subscript), is calculated as $\boldsymbol{\varepsilon}_{**1,p} = (y_{001}-y_{000}+y_{011}-y_{010}+y_{101}-y_{100})/3$.  Compared to the full background-averaged epistasis, the partial averages just leaves out the last term, $y_{111}-y_{110}$, which represents the unavailable phenotype of the triple mutant $y_{111}$.
More generally, we can define this measure of epistasis as another special case of the Hadamard transform:

\begin{equation}
\boldsymbol{\bar{\varepsilon}}_p = \boldsymbol{W}_{_{\!p}} \left( \boldsymbol{Z}_{_{p}}\! \circ \boldsymbol{H} \right) \boldsymbol{\bar{y}}, \label{eq:PartBackg}
\end{equation}

\noindent where $\circ$ designates the element-wise product. $\boldsymbol{W}_{_{\!p}}$ is again a diagonal weighting vector, now given by $v_{ii} = (-1)^{q_i}/T_{p,q_i}$ where $q_i$ is the epistatic order associated with row $i$ as defined earlier, and $T_{p,q_i} = \sum^{p-q_i}_{j=0}{{n-q_i}\choose{j}}$.  Note that $p\ge{}q_i$ because mutants of order higher than $p$ are considered absent in the dataset.

\begin{figure}[t]
\vspace*{-.75cm}
\centerline{\includegraphics[width=.45\textwidth]{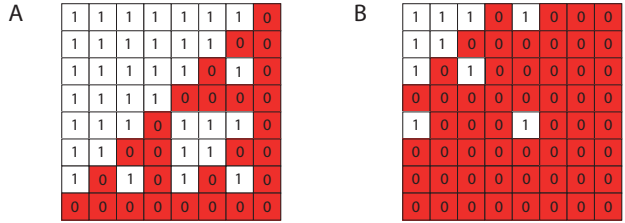}}
\vspace*{.5cm}
\caption{Examples of matrices $\boldsymbol{Z}_{_{p}}$ introduced to calculate the partial background-averaged epistasis, for $n = 3$.
(A) $\boldsymbol{Z}_{_{2}}$ for when data for mutants up to second-order is available and (B) $\boldsymbol{Z}_{_{1}}$ for when only first-order mutants are available. Both matrices are self-similar, which allows their generation for arbitrary order, and are related to the so-called logical Sierpiski triangle. For example $\boldsymbol{Z}_{_{2}} = 1 - \boldsymbol{A} \boldsymbol{\Sigma}$, where $\boldsymbol{A}$ is the anti-diagonal identity matrix and $\boldsymbol{\Sigma}$ is the Sierpinski matrix (i.e. multigrade AND in Boolean logic) for three inputs.\label{fig:Sierpinski}}
\vspace*{-.05cm}
\end{figure}

The matrix $\boldsymbol{Z}_{_{p}}$ simply serves to multiply by zero the terms in the Hadamard matrix that include orders higher than $p$.  Interestingly, the $\boldsymbol{Z}_{_{p}}$ matrices display a self-similar hierarchical pattern (Fig. \ref{fig:Sierpinski}) and are related to so-called Sierpinski triangles (see ref \cite{CRScAcParis160_302}). This permits a recursive definition in both $n$ and $p$ for the product $\boldsymbol{Z}_{_{p}}\! \circ \boldsymbol{H}$, which we will designate as $\boldsymbol{F}_{\!_{n,p}}$:
\begin{equation}
\boldsymbol{F}_{\!_{n,p}}
=
\begin{pmatrix*}[r]
\ \boldsymbol{F}_{\!_{n-1,p}} \ \ \ & \boldsymbol{F}_{\!_{n-1,p-1}}\  \\[0.2em]
\ \boldsymbol{F}_{\!_{n-1,p-1}} & -\boldsymbol{F}_{\!_{n-1,p-1}}\
\end{pmatrix*} \label{eq:PartBackgIter}
\end{equation}\ \\
with $\boldsymbol{F}_{\!_{n,p}}\! = \boldsymbol{H}_{\!_{n}}$ for $n\le{p}$, and $\boldsymbol{F}_{\!_{n,0}}$ is a $2^n \times 2^n$ matrix of zeros, except for a 1 in the upper left corner.  This analysis assumes that data are complete up to the order $p$.  If not, analytical schemes for background-averaged epistasis such as Eqs\,\ref{eq:PartBackg}-\ref{eq:PartBackgIter} are not obvious.

A second analytically tractable case for incomplete data arises in regression, where the idea is to estimate epistatic terms up to a specified order from available data. This involves solving a set of equations similar to the normal equations:
\begin{equation}
\boldsymbol{\tilde{\beta}} = \boldsymbol{Q} \left(\boldsymbol{\tilde{X}}^T \boldsymbol{\tilde{X}}\right)^{-1} \boldsymbol{\tilde{X}}^T \boldsymbol{M} \ \boldsymbol{\bar{y}}
\label{eq:MissY}
\end{equation}
where $\boldsymbol{M}$ is an $s \times 2^n$ matrix constructed from the $2^n$ by $2^n$ identity matrix by deleting the $2^n - s$ rows corresponding to the unavailable phenotypic data, and $\boldsymbol{\tilde{X}} = \boldsymbol{M}  \boldsymbol{X} \boldsymbol{Q}$, with $\boldsymbol{Q}$ defined as above. In order for this system of equations to be solvable, a necessary constraint is that $s\ge{m}$; that is, the number of data points available should be larger than or equal to the number of regression parameters. In addition, the data must be such that it is possible to uniquely solve for all epistatic terms in the regression. For example, if two mutations always co-occur in the data, it is obviously impossible to calculate their independent effects. In such cases, the number of solutions to Eq. \refeq{eq:MissY} is infinite ($\boldsymbol{\tilde{X}}^T \boldsymbol{\tilde{X}}$ is not invertible).

In practice, even with "high-throughput" assays, we can only hope to measure a tiny fraction of all combinatorial mutants due to the vast number of possibilities. In this situation, the problem of inferring epistasis by regression may be further constrained by imposing additional conditions, termed regularization.
For example, kernel ridge regression \cite{ElementsStatLearn} and LASSO \cite{JRStatSocB71_1467} include a weighted norm of the regression coefficients in the minimization procedure. Regularization comes with its own set of caveats \cite{OtwinowskiPNAS}, but its application is, unlike the approaches in Eq. \refeq{eq:PartBackg} and \refeq{eq:MissY}, not conditional on specific structure of the data or depth of coverage.

However, none of these approaches directly addresses the problem of optimally defining appropriate ensembles of genotypes over which averages should be taken. In principle, the idea should be to perform background averaging over a representative ensemble of systems that show invariance of functional properties of interest. How can we generally find such ensembles without the impractical notion of exhaustive functional analysis of the space of possible genotypes?  One idea is motivated by the empirical finding of \emph{sparsity} in the pattern of key epistatic interactions within biological systems.  Indeed, evidence suggests that in proteins, the architecture is to have a small subset of amino acids that shows strong and distributed epistatic couplings surrounded by a majority of amino acids that are more weakly and locally coupled \cite{RanganathanCell,SadovskyPNAS,Kay1,Kay2,FreireAnnRevBBS}. More generally, the notion of a sparse core of strong couplings surrounded by a milieu of weak couplings has been argued to be a signature of evolvable systems \cite{KirschnerPNAS}. If it can be more generally verified, the notion of sparsity might be exploited to define relevant strategies for optimally learning the epistatic structure of natural systems. One approach is to minimize the so-called $\ell_1$-norm (the sum of absolute values of the epistatic coefficients) in a constrained optimization, which has the effect of producing many epistatic coefficients with zero or very small values \cite{JRStatSocB71_1467}, while projecting onto background-averaged epistatic terms:
\begin{equation}
\min_{\boldsymbol{\bar{\varepsilon}}} ||\boldsymbol{\bar{\varepsilon}}||_1 \mathrm{\ \ subject\ to\ \ } \boldsymbol{\bar{y}} = \boldsymbol{H}^{-1} \boldsymbol{V}^{-1} \boldsymbol{\bar{\varepsilon}} \label{eq:CS}
\end{equation}
This procedure is akin to the technique of compressed sensing \cite{CompressedSensing}, a powerful approach used in signal processing to recognize the low-dimensional space in which the relevant features of a high-dimensional dataset occur given the assumption of sparsity of these features. The application of this theory for mapping biological epistasis has to our knowledge not been reported before, but might be explored with focused high-order mutational analyses in specific well-chosen model systems. The necessary technologies for such experiments are now becoming available, and should help define practical data collection strategies for studying epistasis more generally.

It is worth pointing out that other approaches that use ensemble-averaged information to understand biological systems have been developed and experimentally tested. For example, statistical methods that operate on multiple sequence alignments of proteins \cite{MarksNatBiotechnol,RanganathanCell} calculate quantities related to epistasis that are averaged over the space of homologous sequences.  Importantly, these approaches have been successful at revealing a hierarchy of cooperative interactions between amino acids that range from local structural contacts in protein tertiary structures \cite{Weigt,Skerker} to more global functional modes \cite{RanganathanCell}.  For defining good experimental approaches to epistasis, a conceptual advance may come from an attempt to formally map the constrained optimization problem described in Eq.\,\ref{eq:CS} to the kind of ensemble averaging that underlies the statistical coevolution approaches.

\section{Discussion}
A fundamental problem is to define the epistatic structure of biological systems, which holds the key to understanding how phenotype arises from genotype. Here we provide a unified mathematical foundation for epistasis in which different approaches are found to be versions of a single mathematical formalism - the weighted Walsh-Hadamard transform. In the most general case, this transform corresponds to an averaging of mutant effects over all possible genetic backgrounds at every order order of epistasis.  This approach corrects the effect of mutations at every level of epistasis for higher order terms.  Importantly, it represents the degree to which the effects of mutations are transferable from one model system to another, the usual purpose of most mutagenesis studies.  In contrast, the thermodynamic mutant cycle \cite{JMolBiol214_613} (commonly used in biochemistry) constitutes a special case of taking a single reference genotype and thus no averaging \cite{Zaremba, Fuentes, SadovskyPNAS,Yifrach2,Hidalgo,Carter,Ranganathan2}.  This analysis represents the effects of mutations that are specific to a particular model system.  Regression (commonly used in evolutionary biology) is an attempt to capture features of a system with epistatic terms up to a defined lower order, often to bound the extent of epistasis or to predict the effects of higher-order combinations of mutations \cite{Bonhoeffer}.  The similarity of the regression operator to that of the mutant cycle (see Eq. \refeq{eq:recursiveX}) indicates that this approach is also focused around the local mutational environment of a chosen reference sequence.

In general, background averaging would seem to provide the most informative representation of the effect of a mutation.  However, with the exception of very small-scale studies focused in the local mutational environment of extant systems, it is both impractical and logically flawed to collect combinatorially complete mutation datasets for any system.  Thus, the essence of the problem is to define optimal strategies for collecting data on ensembles of genotypes that is sufficient for discovering the biologically relevant epistatic structure of systems.

The notion of sparsity in epistasis provides a general basis for developing such a strategy, and it will be interesting to test practical applications of this concept (e.g. Eq.\,\ref{eq:CS}) in future work.  Defining optimal data collection strategies will not only provide practical tools to probe specific systems, but might guide us to principles underlying the "design" of these systems through the process of evolution, and help the rational design of new systems. The mathematical relations discussed here provide a necessary foundation to advance such understanding.

\begin{acknowledgments}
We thank E. Toprak, K. Reynolds, and members of the Ranganathan laboratory for critically reading the manuscript. FJP gratefully acknowledges funding by the Helen Hay Whitney Foundation sponsored by the Howard Hughes Medical Institute. RR acknowledges support from the Robert A. Welch Foundation (I-1366, R.R.) and the Green Center for Systems Biology.
\end{acknowledgments}

\end{article}

\section*{Supplementary Information: Proofs and extended methods}
\makeatletter
\renewcommand{\thefigure}{S\@arabic\c@figure}
\makeatother
\setcounter{figure}{0}
\renewcommand*{\thefootnote}{\fnsymbol{footnote}}
\setcounter{footnote}{3}\ \\
\subsubsection*{A. Expressing the biochemical epistasis operator $\boldsymbol{G}$ as a Hadamard transform:}\ \\
$\boldsymbol{G} = \boldsymbol{X}^{^{-1}} = \boldsymbol{V} \boldsymbol{X}^T\! \boldsymbol{H}$ \ \ \ (Eq. \refeq{eq:DeltaGHadamard})\\\\\\
First we write the different matrix operators in their recursive form, and then proceed by induction.
We have for the recursive form of $\boldsymbol{X}$:\\\\
$\boldsymbol{X}_{n+1}
=
\begin{pmatrix*}[r]
\ \boldsymbol{X}_n &  \ 0 \ \ \,    \\[0.2em]
\ \boldsymbol{X}_n & \boldsymbol{X}_n\,
\end{pmatrix*}\ \ \mathrm{with}\ \ \
\boldsymbol{X}_0 = 1$\\\\
\\
In order to find the generative function for the inverse $\boldsymbol{X}^{^{-1}}$ we can write $\boldsymbol{X}_{n + 1} \boldsymbol{X}^{^{-1}}_{\ n + 1} = \mathbb{I}$:\\\\
$\begin{pmatrix*}[r]
\ \boldsymbol{X}_n &  \ 0 \ \ \,    \\[0.2em]
\ \boldsymbol{X}_n & \boldsymbol{X}_n\,
\end{pmatrix*} \boldsymbol{X}^{^{-1}}_{n + 1} =
\begin{pmatrix*}[r]
\ \ \mathbb{I} & \ \ 0 \ \ \\[0.2em]
\ \ 0 & \ \ \mathbb{I}\ \
\end{pmatrix*}$,\\\\
which we can solve by Gauss-Jordan elimination:\\\\
$\left(
\begin{array}{rr|rr}
\boldsymbol{X}_n & \ \ 0 \ \ \  & \ \ \mathbb{I} & \ \ 0 \\[0.2em]
\boldsymbol{X}_n & \boldsymbol{X}_n\ & \ 0 & \ \ \mathbb{I}\\
\end{array}
\right) \Rightarrow
\left(
\begin{array}{rr|rr}
\mathbb{I}& \ \ 0 \ \  & \boldsymbol{X}^{^{-1}}_{\ n} & 0 \ \ \ \\[0.2em]
\mathbb{I} & \mathbb{I} \ \ & 0 \ \ \ & \boldsymbol{X}^{^{-1}}_{\ n}\\
\end{array}
\right)\Rightarrow
\left(
\begin{array}{rr|rr}
\mathbb{I}& \ \ 0 \ \  & \boldsymbol{X}^{^{-1}}_{\ n} & 0 \ \ \ \\[0.2em]
0 & \mathbb{I} \ \ & -\boldsymbol{X}^{^{-1}}_{\ n} & \boldsymbol{X}^{^{-1}}_{\ n}\\
\end{array}
\right)
$\\\\
hence we have for the inverse of $\boldsymbol{X}$:\\\\
$\boldsymbol{X}^{^{-1}}_{n+1}
=
\begin{pmatrix*}[r]
\ \boldsymbol{X}^{^{-1}}_{\ n} & \ \ 0 \ \ \ \  \\[0.2em]
\ -\boldsymbol{X}^{^{-1}}_{\ n} & \boldsymbol{X}^{^{-1}}_{\ n} \
\end{pmatrix*}\ \ \mathrm{with}\ \ \
\boldsymbol{X}^{^{-1}}_0 = 1$\\\\
\\
Which is identical to the recursive form for $\boldsymbol{G}$:\\\\
$\boldsymbol{G}_{n+1}
=
\begin{pmatrix*}[r]
\ \boldsymbol{G}_n & \ \ 0 \ \ \   \\[0.2em]
\ -\boldsymbol{G}_n & \boldsymbol{G}_n\
\end{pmatrix*}$\\\\
\\
We further have:\\\\
$\boldsymbol{H}_{n+1}
=
\begin{pmatrix*}[r]
\ \ \boldsymbol{H}_n & \boldsymbol{H}_n\  \\[0.2em]
\ \ \boldsymbol{H}_n & -\boldsymbol{H}_n\
\end{pmatrix*}\ \ \mathrm{with}\ \ \
\boldsymbol{H}_0 = 1$\\\\
\\
and
$\boldsymbol{V}_{\!n+1}
=
\begin{pmatrix*}[r]
\ \frac{1}{2} \boldsymbol{V}_{\!n} & 0 \ \ \ \\[0.2em]
0 \ \ \ & -\boldsymbol{V}_{\!n}\
\end{pmatrix*}\ \ \mathrm{with}\ \ \
\boldsymbol{V}_{\!0} = 1$\\\\\\
\\
With the above relations we can derive the equality in the main text expressing $\boldsymbol{G}$ as a Hadamard transform:\\\\
$\boldsymbol{G}_{_{n}} = \boldsymbol{X}_{_{n}}^{^{-1}} = \boldsymbol{V}_{_{\!n}} \boldsymbol{X}_{_{n}}^T \boldsymbol{H}_{_{n}}$\\\\
For $n = 0$ the statement is trivial. We now show by induction that this relation holds for all $n$.\\\\
$\boldsymbol{G}_{_{n+1}}
=
\begin{pmatrix*}[r]
\ \boldsymbol{G}_n & \ \ 0 \ \ \  \\[0.2em]
\ -\boldsymbol{G}_n & \boldsymbol{G}_n \
\end{pmatrix*}
=
\begin{pmatrix*}[r]
\  \boldsymbol{V}_{_{\!n}} \boldsymbol{X}_{_{n}}^T \boldsymbol{H}_n & 0 \ \ \ \ \ \ \\[0.2em]
\  -\boldsymbol{V}_{_{\!n}} \boldsymbol{X}_{_{n}}^T \boldsymbol{H}_n   &  \boldsymbol{V}_{_{\!n}} \boldsymbol{X}_{_{n}}^T \boldsymbol{H}_n\
\end{pmatrix*}
$\\\\\\
$
\hspace*{.85cm}=
\begin{pmatrix*}[r]
\ \frac{1}{2} \boldsymbol{V}_{\!n} & 0 \ \ \ \\[0.2em]
0 \ \ \ & -\boldsymbol{V}_{\!n}
\end{pmatrix*}
\begin{pmatrix*}[r]
\ \ 2 \boldsymbol{X}_{_{n}}^T \boldsymbol{H}_n & 0 \ \ \ \ \ \\[0.2em]
\ \ \boldsymbol{X}_{_{n}}^T \boldsymbol{H}_n\  & -\boldsymbol{X}_{_{n}}^T \boldsymbol{H}_n\
\end{pmatrix*}
=
\begin{pmatrix*}[r]
\ \frac{1}{2} \boldsymbol{V}_{\!n} & 0 \ \ \ \\[0.2em]
0 \ \ \ & -\boldsymbol{V}_{\!n}\
\end{pmatrix*}
\begin{pmatrix*}[r]
\ \boldsymbol{X}_{_{n}}^T & \boldsymbol{X}_{_{n}}^T\ \   \\[0.2em]
\ 0 \ \  & \boldsymbol{X}_{_{n}}^T\ \
\end{pmatrix*}
\begin{pmatrix*}[r]
\ \ \boldsymbol{H}_n & \boldsymbol{H}_n\  \\[0.2em]
\ \ \boldsymbol{H}_n & -\boldsymbol{H}_n\
\end{pmatrix*}$\\\\\\[0.7em]
$
\hspace*{.85cm}=
\boldsymbol{V}_{_{\!n+1}} \boldsymbol{X}_{_{n+1}}^T \boldsymbol{H}_{_{n+1}}\ \ \ \ \ \ \  \ \mathrm{QED}$\\\\\\
\\

\subsubsection*{B. Expressing the regression operator as a Hadamard transform:}\ \\
$\boldsymbol{Q} \left(\boldsymbol{\hat{X}}^T\, \boldsymbol{\hat{X}}\right)^{-1}\! \boldsymbol{\hat{X}}^T = \boldsymbol{V} \boldsymbol{X}^T \boldsymbol{S}  \boldsymbol{H}$ \ \ \ (Eq. \refeq{eq:regressHadamard})\\\\\\
We will use $\boldsymbol{\hat{X}} = \boldsymbol{X} \boldsymbol{Q}$ and $\boldsymbol{S} = \boldsymbol{Q} \boldsymbol{Q}^T$ as defined in the main text.\\\\
For the right-hand side we can write\\\\
$\boldsymbol{V} \boldsymbol{X}^T \boldsymbol{S} \boldsymbol{H} = \frac{1}{2^n} \boldsymbol{V} \boldsymbol{X}^T \left(\boldsymbol{H} \boldsymbol{H}\right) \boldsymbol{S} \boldsymbol{H}$\\\\
where we used $\boldsymbol{H}_{_{n}}^2 = 2^n \mathbb{I}_{_{n}}$, which can be proven straightforwardly by induction using the generative function for $\boldsymbol{H}$.\\
Rearranging and using $\boldsymbol{X}^{^{-1}} = \boldsymbol{V} \boldsymbol{X}^T \boldsymbol{H}$, we obtain\\\\
$\boldsymbol{V} \boldsymbol{X}^T \boldsymbol{S} \boldsymbol{H} = \frac{1}{2^n} \boldsymbol{X}^{^{-1}} \left(\boldsymbol{H} \boldsymbol{S} \boldsymbol{H} \right)$\\\\
We thus have to prove\\\\
$\boldsymbol{Q} \left(\boldsymbol{\hat{X}}^T\, \boldsymbol{\hat{X}}^{}\right)^{-1}\! \boldsymbol{\hat{X}}^T = \frac{1}{2^n} \boldsymbol{X}^{-1} \left(\boldsymbol{H} \boldsymbol{S} \boldsymbol{H} \right)$\\\\
Left-multiplying both sides by $\boldsymbol{\hat{X}}^T \boldsymbol{X}$ (mind the hat is only on the first operator) and right-multiplying by $\boldsymbol{H}$ we are left to prove\\\\
$\boldsymbol{\hat{X}}^T \boldsymbol{H} = \boldsymbol{\hat{X}}^T \boldsymbol{H} \boldsymbol{S}$\\\\
Left-multiplication by $\boldsymbol{Q}$ yields\\\\
$\boldsymbol{S} \boldsymbol{X}^T \boldsymbol{H} = \boldsymbol{S} \boldsymbol{X}^T \boldsymbol{H} \boldsymbol{S}$\\\\
which, again using the relation we proved in section A above, can be rewritten as\\\\
$\boldsymbol{S}  \boldsymbol{V}^{-1} \boldsymbol{X}^{-1} = \boldsymbol{S} \boldsymbol{V}^{-1} \boldsymbol{X}^{-1} \boldsymbol{S}$\\\\
or\\\\
$\boldsymbol{S} \boldsymbol{X}^{-1} = \boldsymbol{S} \boldsymbol{X}^{-1} \boldsymbol{S}$\\\\
given the commutative properties of diagonal matrices $\boldsymbol{S}$ and $\boldsymbol{V}^{-1}$.\\
\\
This equality indicates that setting certain rows of $\boldsymbol{X}^{-1}$ to zero (left-hand side) is the same as setting both those rows and corresponding columns of $\boldsymbol{X}^{-1}$ to zero (right-hand side). This is obviously not true for every set of rows and columns, and needs more discussion.\\
\\
We can prove this iteratively starting at regression to order $n-1$ and going down to lower order. If regression is done to order $n-1$, this means that only the last row of $\boldsymbol{X}^{-1}$ is set to zero, and by construction of $\boldsymbol{X}^{-1}$ (see above) the last column only has a non-zero element in this row. This means that in this case the equality is correct. Another way to see this is looking at matrix $\boldsymbol{G}$ for $n=3$ in its explicit representation in the main text (here $\boldsymbol{G}$ being identified with $\boldsymbol{X}^{-1}$) and noting that the highest order epistatic term $\gamma_{111}$ is the only one that receives a contribution from the highest order ($n$) mutant term $y_{111}$.\\
\\
Next, if regression is performed instead to order $n-2$, not only the last row of $\boldsymbol{X}^{-1}$ is set to zero, but also the rows corresponding to $n-1$ order mutants. Analogously to above, the only terms in the vector $\boldsymbol{\bar{\gamma}}$ that receive contributions from the $n-1$ order mutants are the ones in the rows corresponding to $n-1$ order of epistasis (since the row corresponding to $n$\sups{th} order is already set to zero), meaning that their corresponding column again has only one non-zero element. Hence setting these rows to zero will directly set their corresponding column to zero, and the equality holds.\\
\\
And so forth for regression to order $n-3$, etc., etc.\\
\\
QED\\\\

\end{document}